\documentclass{aa}
\usepackage{graphicx}
\usepackage{txfonts}

\begin{document}

\title{Application of fast CCD drift scanning to speckle imaging of binary
stars}
\titlerunning{Speckle imaging with fast CCD drift scanning}

\author{O. Fors\inst{1}\fnmsep\inst{2}
\and E.P. Horch\inst{3}
\and J. N\'u\~nez\inst{1}\fnmsep\inst{2}}

\offprints{O. Fors, \email{ofors@am.ub.es}}

\institute{Departament d'Astronomia i Meteorologia, Universitat de  Barcelona,
Av. Diagonal 647, 08028 Barcelona, Spain
\and
Observatori Fabra, Cam\'{\i} de l'Observatori s/n, 08035 Barcelona, Spain
\and
Department of Physics, University of Massachusetts Dartmouth, 285 Old Westport
Road, North Dartmouth, MA, USA
}

\date{Received / Accepted}

%________________________________________________________________
\abstract{

A new application of a fast CCD drift scanning technique that allows us to
perform speckle imaging of binary stars is presented. For each
observation, an arbitrary number of speckle frames is periodically stored
on a computer disk, each with an appropriate exposure time given both
atmospheric and instrumental considerations. The CCD charge is shifted
towards the serial register and read out sufficiently rapidly to avoid an
excessive amount of interframe dead time. Four well-known binary systems
(\object{ADS 755}, \object{ADS 2616}, \object{ADS 3711} and \object{ADS
16836}) are observed in to show the feasibility of the proposed technique.

Bispectral data analysis and power spectrum fitting is carried out for each
observation, yielding relative astrometry and photometry. A new approach for
self-calibrating this analysis is also presented and validated.

The proposed scheme does not require any additional electronic or optical
hardware, so it should allow most small professional observatories and advanced
amateurs to enjoy the benefits of diffraction-limited imaging.

%________________________________________________________________
}

\maketitle

%________________________________________________________________
\section{Introduction} \label{introduction}

Over the last few years, CCDs have been used with increasing frequency for
speckle imaging. This work actually started more than a decade ago with
the work of Beletic \& Zadnik (\cite{zadnik93}), and has more recently
been extended by Horch et al. (\cite{horch97a}, \cite{horch99}) and
Kluckers et al (\cite{kluckers97}).

There are three main reasons for this change. First, CCDs have
dramatically improved in terms of their readout noise and speed
characteristics. Second, it has been realized that changing the readout
pattern allows one to use large-format CCDs effectively in speckle
imaging. Finally, there has been the hope that CCDs would allow
diffraction-limited photometry in a way that intensified cameras, such as
ICCDs and other microchannel--plate-based devices have not been able to do
up to the present.

In this paper, we show that CCD-based speckle imaging can be extended to a
relatively inexpensive detection system, namely the Santa Barbara Instruments
Group (SBIG) ST-8 camera, which in particular does not have the speed and
readout flexibility that other CCDs being used for the same purpose have.

Fast drift scanning has already been shown to be a useful technique for
recording lunar occultations at millisecond sampling rate (Fors et al.
\cite{fors01}, \cite{fors03}).

In the case of speckle observations, a fast drift scanning acquisition
scheme is adapted to meet the specific needs of speckle imaging. The data
are read out off the chip as fast as possible, but with pauses in readout
corresponding to the collection of speckle patterns. In addition, we
introduce a method of self-calibration of speckle data in the Fourier
plane which makes possible to make reconstructed images without taking
data on an unresolved point source. Finally, we make some recommendations
which may be valuable for small observatories or advanced amateurs should
they wish to carry this work forward.

%________________________________________________________________
\section{Proposed technique basis} \label{technique}

\begin{figure*}
\resizebox{\hsize}{!}{\includegraphics{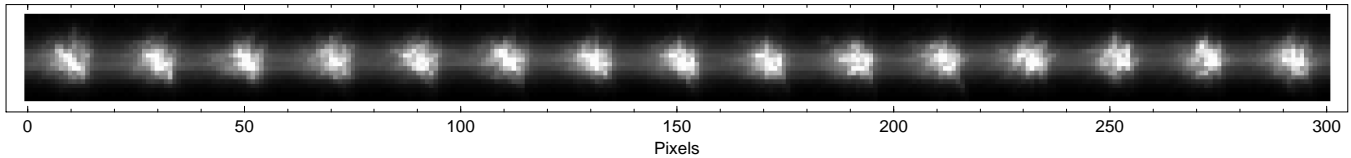}}
\caption{Raw strip image of \object{ADS755} as observed when following the proposed technique.
Specklegrams are 20x20 pixels in size and the exposure time is 39ms.}
\label{strip}
\end{figure*}

Fors et al. (\cite{fors01}, \cite{fors03}) showed that fast drift scanning
can be applied to obtain high-resolution measurements by means of lunar
occultation (LO) observations. In that approach, the occultation
lightcurve was recorded by reading out every millisecond the small
fragment of the CCD in which the object was situated. This procedure was
continuously maintained until the occultation event took place.

In this paper we present a variation of the former acquisition technique.
As in the LO approach, the telescope tracking is turned on and the shutter
remains open throughout the observation. To perform speckle imaging, the
continuous column readout is periodically interrupted by an amount of time
which is the effective speckle frame exposure time. The resulting image of
this process is an arbitrarily long strip with a series of speckle frames.

Of course, the camera spends some measurable time reading out all columns of
each speckle frame. As a result of that unavoidable dead time between
consecutive speckle frames, a low-level streaking appears between speckle
images. In general, the importance of this effect will depend on the camera
specifications, namely digitization and data transfer rate. 

The proposed acquisition scheme is directly applicable to any full frame
CCD camera for which it is possible to set readout column rate and size by
software means. No hardware or optical modification has to be made to the
telescope to make this technique possible.

Large format CCDs had already been used for speckle imaging in the past by
one of us (Horch et al. \cite{horch97a}). In that approach, called fast
subarray readout, ten to twenty speckle frames were stored in a subarray
strip of the KAF-4200 chip until it became filled. Afterwards, the shutter
was closed and the whole subarray was readout. The proposed technique in
the current article exhibits one advantage and one disadvantage with
respect to fast subarray readout. On the one hand, one can now obtain as
many speckle frames as desired without periodically closing the shutter:
it is not limited by CCD chip size as in the subarray-readout mode. On the
other hand, it is necessary to read out all the columns of the CCD between
consecutive speckle frame exposures. This yields a longer dead time, which
increases low-level streaking. However, it is likely the dead time will be
significantly reduced in the very near future with new faster CCD cameras
available on the professional and high-end amateur market (see
Sect.~\ref{improv} for further discussion on this topic).

The term {\it fast drift scanning} for both speckle imaging and lunar
occultations may seem somewhat ambiguous. In a strict sense, term should
only be used when the R.A. tracking drive is turned off and, as a result,
the imaged scene {\it drifts} over the CCD chip at the same rate the
column charge is clocked towards the serial register. However, to be
consistent with Fors et al. (\cite{fors01}) we will adopt the same
designation.

%________________________________________________________________
\section{Observations} \label{obs}

Speckle observations were conducted at the 1.5\,m telescope of the
Observatorio Astronomico Nacional at Calar Alto (Spain) in October, 2001.
The same camera employed by LO at Fors et al.
(\cite{fors01},\cite{fors03}) was used. This is a Texas Instruments TC-211
CCD, set inside an SBIG ST8 camera as the tracking chip. It is a full
frame front-illuminated CCD with 13.75x16 ${\mu}$m pixels and a 192x164
pixel format. It is read out through a parallel port interface and its
electronic module can be operated at 30 kHz with 12 electrons rms readout
noise. The camera was directly attached at the Cassegrain focus of the
telescope without any magnification optics. This configuration yields an
effective focal length of 12280mm.

Four binary systems were observed during 5 nights (see Cols.~1--6 in
Table~\ref{measures} for details), under median seeing conditions of
$1.3^{\prime \prime}$. Those objects were selected because they have well
determined orbits which allow us to validate the acquisition technique
described in Sect.~\ref{technique}. Several speckle frame sequences were
obtained for every object. Each sequence consists of a few hundreds of frames.

All speckle observations were conducted with a Cousins R filter
($\lambda=641\pm100\,{\rm nm}$). At this wavelength, the  diffraction-limited spot
size is equal to $108\,{\rm mas}$. On the other hand, the scale calibration was
carried out by means of a standard plate solution of long exposure frames, and
was found to be $9.375\,{\rm mas}\,{\rm mm^{-1}}$. Thus, our data is undersampled and this
will be taken into account in the reduction process (see Sect~\ref{analysis}).

Data acquisition was performed using an implementation of the proposed
technique into a DOS-based program called SCAN\footnote{available at {\tt
http://www.driftscan.com}}. This program offers satisfactory relative timing
accuracy when scheduling column readout at millisecond rates. 

In CCD-based speckle imaging there is a competition between readout noise and
atmospheric correlation time. On the one hand, longer frame integration times
give you more photons, which gives better contrast of the speckle pattern with
the readout noise. On the other hand, you lose speckle contrast if too long a
frame time is used. Therefore, it is not just an  instrumental readout
limitation that forces us to use a frame time longer than the correlation time,
but it is desirable to minimize the effect of CCD read noise.

In Fig.~\ref{strip} we show a subset of a typical sequence of speckle frames
obtained by means of this technique. For this particular case, a 20-pixel
column is stored every 1.8\,ms on average, yielding a dead time of 36\,ms. This
must be added to the exposure time, 39\,ms. Note that this is significantly
larger than the typical atmospheric coherence time for seeing of 1.3
arcseconds, which has been estimated at several observatories to be on the
order of 4-8\,ms. The choice of this longer exposure time and its consequences
for data quality is justified and discussed in Sect.~\ref{analysis}.

%_______________________________________________________________
\section{Data Analysis and Self-Calibration} \label{analysis}

Once the raw data are read out of the camera, pixels around the object of 
interest are extracted and converted to FITS format. The FITS file  is stored
as an image stack where each image contains a 20 x 20 pixel speckle pattern.
Approximately 500 of such images are contained in the  stack of a single
observation. These files are then analyzed in exactly the same way as described
in Horch et al. (\cite{horch97a}). Briefly, the method is to subtract the bias
level and the streak between images caused by the readout scheme, and then to
compute the autocorrelation and low-order bispectral subplanes needed for
subsequent analysis. 

\begin{figure}
\resizebox{\hsize}{!}{\includegraphics{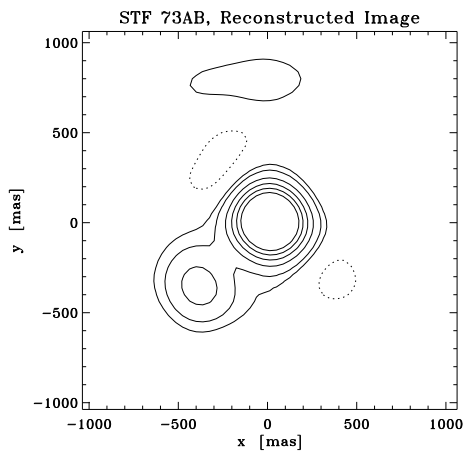}}
\caption{A reconstructed image of WDS 00550+2338 = \object{ADS 755} =  STF 73AB. North is down, East is to the
right. Contours are drawn at -0.05, 0.05, 0.10, 0.20, 0.30, 0.40, and 0.50 of the maximum value in the
array. The dotted contours indicate the value -0.05. The secondary star appears below and to the left
of the primary,  which is located in the center of the image. The feature in the upper part of the
figure is not real and appears to be related to the mismatch between the seeing profile of the binary
observation and the radially generated point source.}
\label{ads755ri}
\end{figure}

In the case of reconstructed images, the relaxation technique of Meng et al.
(\cite{meng90}) is used to generate a phase map of the  object's Fourier
transform, and this is combined with the object's modulus obtained by taking
the square root of the power spectrum. By combining the modulus and the phase
and inverse transforming, one arrives at the reconstructed image. An example of
such an image is shown in Fig.~\ref{ads755ri}. 

In the case of deriving relative astrometry of binary stars, the weighted least
squares approach of Horch et al. (\cite{horch96}) has been used. This method
fits a power spectrum deconvolved by a point source calibrator to a trial 
fringe pattern and then attempts to minimize the reduced chi-squared of the
function. The undersampling correction of Horch et al (\cite{horch97a}) is
applied.

\begin{figure}
\resizebox{\hsize}{!}{\includegraphics{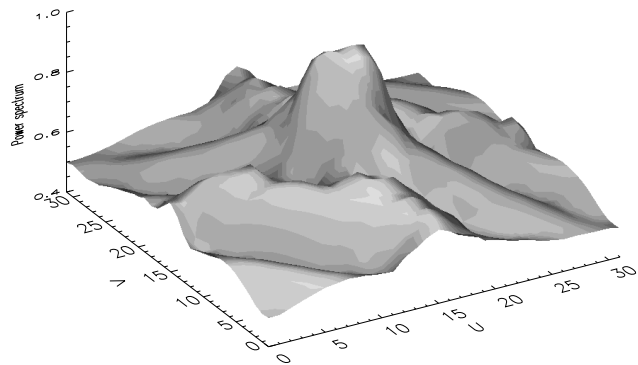}}
\caption{A surface plot of the power spectrum of one of the observing runs for \object{ADS 755}. Note the fringe
pattern due to binarity of the object.}
\label{ads755_pwr}
\end{figure}

\begin{figure}
\resizebox{\hsize}{!}{\includegraphics{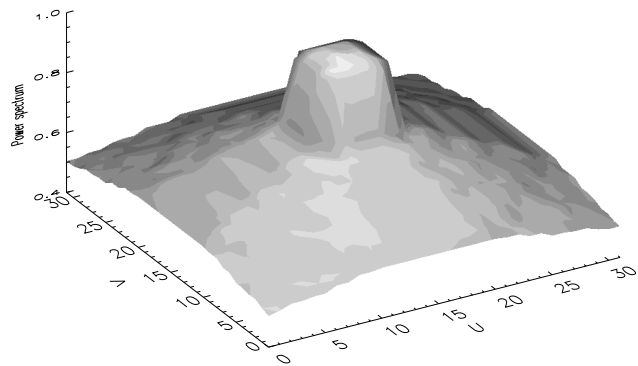}}
\caption{Power spectrum for a calculated point source following the self-calibration scheme. Compared to
Fig.~\ref{ads755_pwr}, the central peak due to seeing remains approximately the same and fringes in the speckle shoulder
are not present, as expected.}
\label{ads755ps_pwr}
\end{figure}

For all data discussed here, an estimate for an unresolved point source power
spectrum was constructed from that of a binary star. This has the advantage of
allowing binary star observations to be taken without interrupting for
measurements of the speckle transfer function. A synthetic point source
estimate can be generated first by forming the power spectrum of any binary
(see Fig.~\ref{ads755_pwr}), and then extracting a trace from the image along
the central fringe. Since the binary is not resolved along this direction, this
is essentially a one-dimensional estimate of an unresolved source. This
one-dimensional function is then rotated about the origin of the frequency
plane to fill a two-dimensional array. This generates a radially symmetric
function, as indeed a true unresolved source should show under perfect conditions
(see Fig.~\ref{ads755ps_pwr}). The method has limitations as we will discuss in
Sect.~\ref{limit} after the main body of results has been presented, but
provides a way to make the deconvolution needed without recourse to point
source observations.

\begin{figure}
\resizebox{\hsize}{!}{\includegraphics{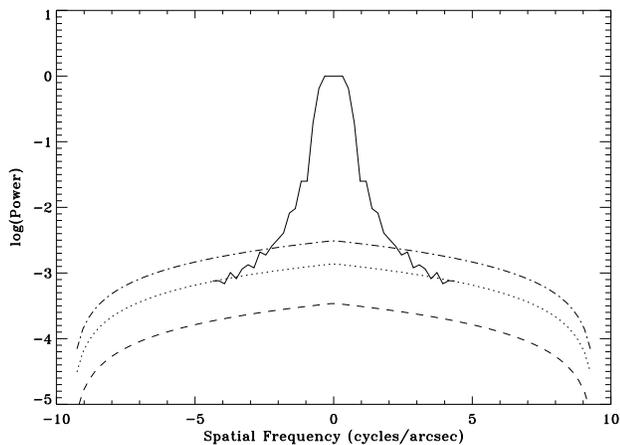}}
\caption{Comparison of cutoff frequencies of observed and simulated 1-D speckle
transfer functions. The former (solid line) was generated from the \object{ADS
2616} point source. The latter represents the diffraction limited power
spectrum obtained at $641\,{\rm nm}$ using a 1.5\,m aperture. Three different
values of the Fried parameter $r_0$, $5\,{\rm cm}$ (dashed), $10\,{\rm cm}$ (dotted), and
$15\,{\rm cm}$ (dash-dotted), have been considered. Note that the better the seeing,
the larger $r_0$ and so the higher the curve on the plot.}
\label{cutoff}
\end{figure}

As noted in Sect.~\ref{obs}, the speckle frame exposure time was chosen to be
larger than the coherence time. This choice is justified by the competition between
readout noise and correlation time when performing CCD-based speckle imaging.
On the one hand, speckle frames show the highest possible signal-to-noise ratio
when the integration time is in fact longer than the coherence time. One of us
(E.H.) has shown that 50\,ms is the exposure time where the maximum in the
signal-to-noise ratio occurs at the WIYN telescope, which uses a CCD with a readout
noise of 10 electrons.  That probably implies a factor of 4 to 5 larger than
the coherence time (Horch et al.~\cite{horch02}). On the other hand, in general
speckle contrast decreases as longer exposure time are used. Therefore, it is
not just an instrumental readout limitation that forces us to use a frame time
longer than the correlation time, but it is desirable to minimize the effect of
CCD read noise, while still preserving sufficient contrast on speckle patterns.

In addition, interframe dead time contributes to low-level streaking.
However, note that the light contributing to streaking is distributed far more
uniformly and over more pixels than those forming the speckle pattern itself. As
a result, the ratio between intensity peaks is much more favorable than the
ratio between dead time and atmospheric coherence time.

All this introduces attenuation in the higher frequencies of our data. To
illustrate how this affects resolution, a plot with four 1-D power
spectrum curves has been made. As shown in Fig.~\ref{cutoff}, one
corresponds to an observed point source and the other three to the
diffraction limited spot one would obtain with the instrumental conditions
of current paper. The attenuation factor used for generating such
simulated profiles is given by:

\begin{equation}
A=0.435 (r_0/D)^2,
\label{attenuation}
\end{equation}

where $r_0$ is the Fried parameter and $D$ the telescope diameter. The
0.435 is a geometrical factor derived by Korff (\cite{korff73}) and Fried
(\cite{fried79}). 

Ideally, the high-frequency portion of the speckle transfer function should
overlap to the simulated curve attenuated with the $r_0$ value which best
matches the real seeing. However, due to the significant undersampling of our
data, the observed power spectrum does not span up to the theoretical
diffraction limit (near $\pm10cycles\,arcsec^{-1}$). It is worth mentioning
that our reduction software does account for the aliasing effect of the
undersampling and, in principle, is able to extract part of those frequencies
which are aliased to lower frequencies. However, this last is somewhat limited
by the low signal-to-noise which these high frequencies show. Thus, we see that
the impact of longer exposure time is relatively small, and does not handicap
our data quality.

%_______________________________________________________________
\section{Results} \label{res}

\begin{table*}
\begin{center}
\caption[]{Double star speckle measures.}
\label{measures}
\begin{tabular}{rlrrccrrl}
\hline \hline
{ADS} & {Discoverer} & HD & HIP & WDS & {Date} & {$\theta$} & {$\rho$} & {$\Delta m$} \\
{} & {Designation} && & {($\alpha$,$\delta$ J2000.0)} & {(BY)} & {($^{\circ}$)} & {($^{\prime \prime}$)}\\
\hline

   755      & STF  73AB              &   5286   &   4288 & $00550+2338$
  & 2001.8127 &   311.5    &  0.935    & 1.17 $^{b}$ \\
&&&&
  & 2001.8178 &   310.5    &  0.936    & 0.43  \\
&&&&
  & 2001.8207 &   311.0    &  0.936    & 0.53  \\
  2616      & STF 412AB              &  22091   &  16664 & $03344+2428$
  & 2001.8208 &  175.8     &  0.651   & 0.35  $^{a}$ \\
&&&&
  & 2001.8261 &  176.2     &  0.646   & 0.58  $^{a}$ \\
  3711      & STT  98                &  33054   &  23879 & $05079+0830$
  & 2001.8157 &   319.8    &  0.743   & 0.52  \\
 16836      & BU  720                & 221673   & 116310 & $23340+3120$
  & 2001.8124 &   95.8    &  0.560     & 0.57 $^{b}$ \\
&&&&
  & 2001.8208 &   97.2    &  0.585     & 0.36  \\
&&&&
  & 2001.8260 &   92.8    &  0.589     & 0.90 $^{b}$ \\
\hline
\end{tabular}
\begin{list}{}{}
\item[a.] There is an ambiguity of 180 degrees in the position angle compared to previous observations.
\item[b.] Observation was taken at low elevation. This may affect the quality of the result.
\end{list}
\end{center}
\end{table*}

\begin{table*}
\begin{center}
\caption[]{Comparison of Results Obtained with Different Point Source Power Spectra.}
\label{comparison}
\begin{tabular}{rlrrccrrl}
\hline \hline
ADS & Discoverer & HD & HIP & WDS & Date & $\theta$ & $\rho$ & $\Delta m$ \\
{}  & Designation &&& ($\alpha$,$\delta$ J2000.0) & (BY) & ($^{\circ}$) & ($^{\prime \prime}$)\\
\hline

   755      & STF  73AB              &   5286   &   4288 & $00550+2338$
  & 2001.8207 &   311.7    &  0.939    & 0.76  \\
&&&&
  & 2001.8207 &   312.5    &  0.945    & 0.98  \\
&&&&
  & 2001.8207 &   311.7    &  0.936    & 0.73  \\
&&&&
  & 2001.8207 &   311.1    &  0.938    & 0.59  \\
&&&& $\sigma$ &&
0.6 & 0.004 & 0.16 \\

&&&&&&&& \\

  2616      & STF 412AB              &  22091   &  16664 & $03344+2428$
  & 2001.8261 &  178.4     &  0.674   & 0.58   \\
&&&&
  & 2001.8261 &  174.5     &  0.642   & 0.31   \\
&&&&
  & 2001.8261 &  176.2     &  0.664   & 0.23   \\
&&&&
  & 2001.8261 &  173.5     &  0.634   & 0.35   \\
&&&&
  & 2001.8261 &  178.4     &  0.672   & 0.70   \\
&&&& $\sigma$ &&
2.2 & 0.018 & 0.20 \\
\hline
\end{tabular}
\end{center}
\end{table*}

\begin{figure*}   
\resizebox{\hsize}{!}{\includegraphics{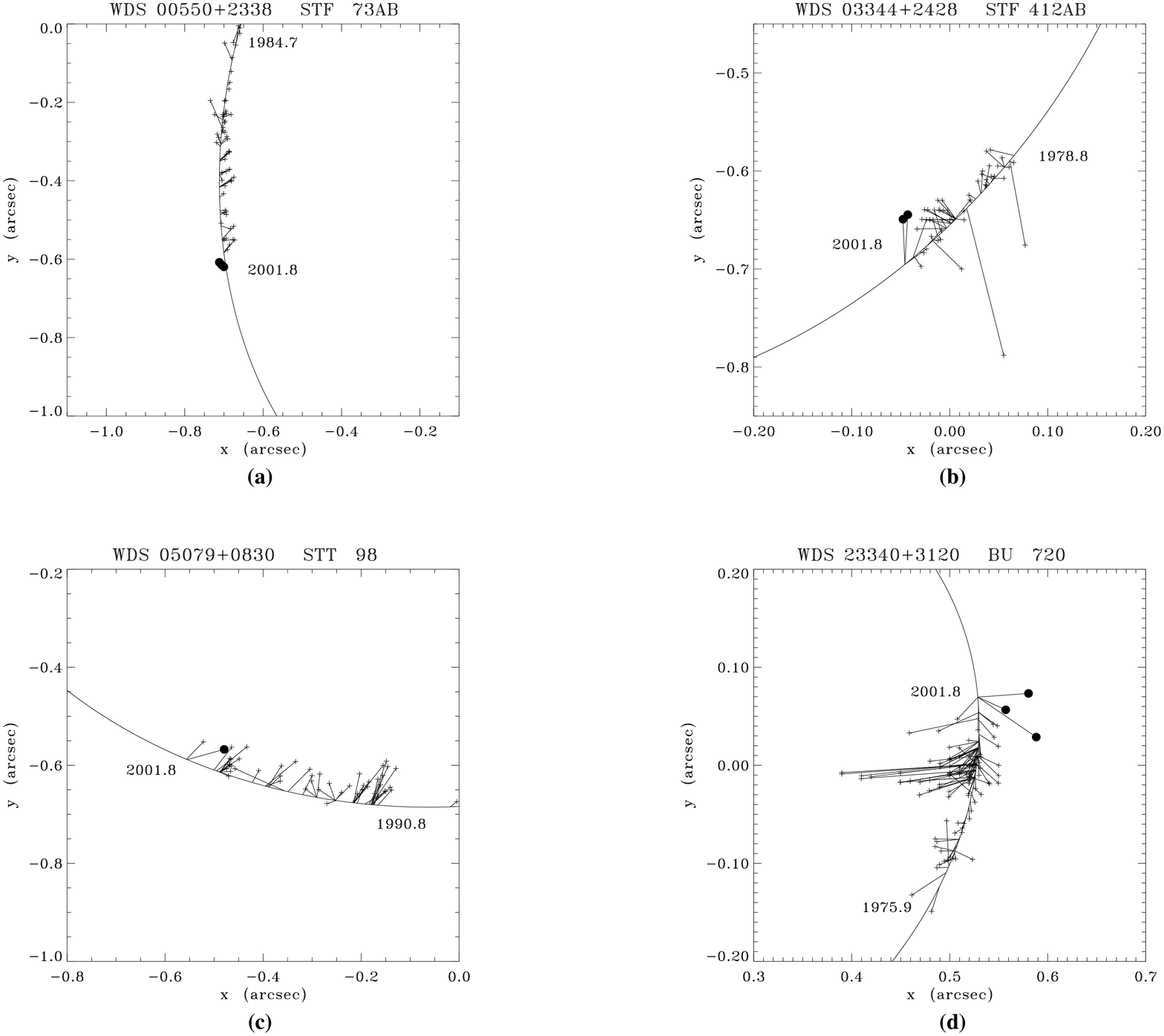}}
\caption{A comparison of the position angle and separation measures presented
here with the work of other observers. In all plots, North is down and East is
to the right. In all cases, the object has an orbit listed in the 6th Catalog
of Orbits of Visual Binary Stars (Hartkopf et al. (\cite{hartkopf03})), and the
orbital trajectory is plotted. Observations of previous observers, compiled by
Hartkopf et al. (\cite{hartkopf02}), are marked with small plus symbols, with
a line segment drawn from the point to the ephemeris prediction for that epoch.
The observations presented here are marked with the solid dots, again with line
segments joining the point to the predicted location given the orbital
elements. (a) WDS 00550+2338 = STF 73AB = \object{ADS 755}. The  orbit plotted is that
of Docobo \& Costa (\cite{docobo90}), rated as a Grade 2 orbit in the Sixth
Catalog. (b) WDS 03344+2428 = STF 412AB = \object{ADS 2616}. The orbit plotted is that
of Scardia et al.\ (\cite{scardia02}), rated as a Grade 3 orbit in the Sixth
Catalog. (c) WDS 05079+0830 = STT 98 = \object{ADS 3711}. The orbit plotted is that of
Baize (\cite{baize69}), rated as a Grade 3 orbit in the Sixth Catalog. (d) WDS
23340+3120 = BU 720 = \object{ADS 16836}. The orbit plotted is that of Starikova
(\cite{starikova82}), rated as a Grade 3 orbit in the Sixth Catalog.} 
\label{orbits}   
\end{figure*}

In Table~\ref{measures} we show all speckle measures obtained during our
five night observing run after applying the self-calibration analysis as
explained in the previous section. Column headings are as follows: (1) the
Aitken Double Star number; (2) the discoverer designation as it appears in
the Washington Double Star Catalog (WDS); (3) the Henry Draper Catalogue
number; (4) the HIPPARCOS Catalogue number; (5) the Washington Double Star
Catalogue number, which is the same as the position in 2000.0 coordinates;
(6) the date in fraction of the Besselian year when the observation was
made; (7) the position angle ($\theta$) with north through east defining
the positive sense of the angle; (8) the separation ($\rho$) in arc
seconds; and (9) the magnitude difference as judged from the speckle
observations. Position angles are not corrected for precession and are
therefore valid for the epoch of observation shown. Every
($\rho$,$\theta$,$\Delta_m$) triplet in the table is the result of
averaging the result of 5 frame sequences, which were exposed within a few
minutes of each other. As indicated in the table, some observations
were taken at low elevation. Note that position angle, separation, and
magnitude differences for these measures appear discrepant from the rest
of values. Therefore, the self-calibration point source method should be used
only at modest zenith angles (less than thirty degrees, if no atmospheric
dispersion compensation is performed). Further discussion about this
limitation will be given in Sect.~\ref{limit}.

\begin{figure*}
\resizebox{\hsize}{!}{\includegraphics{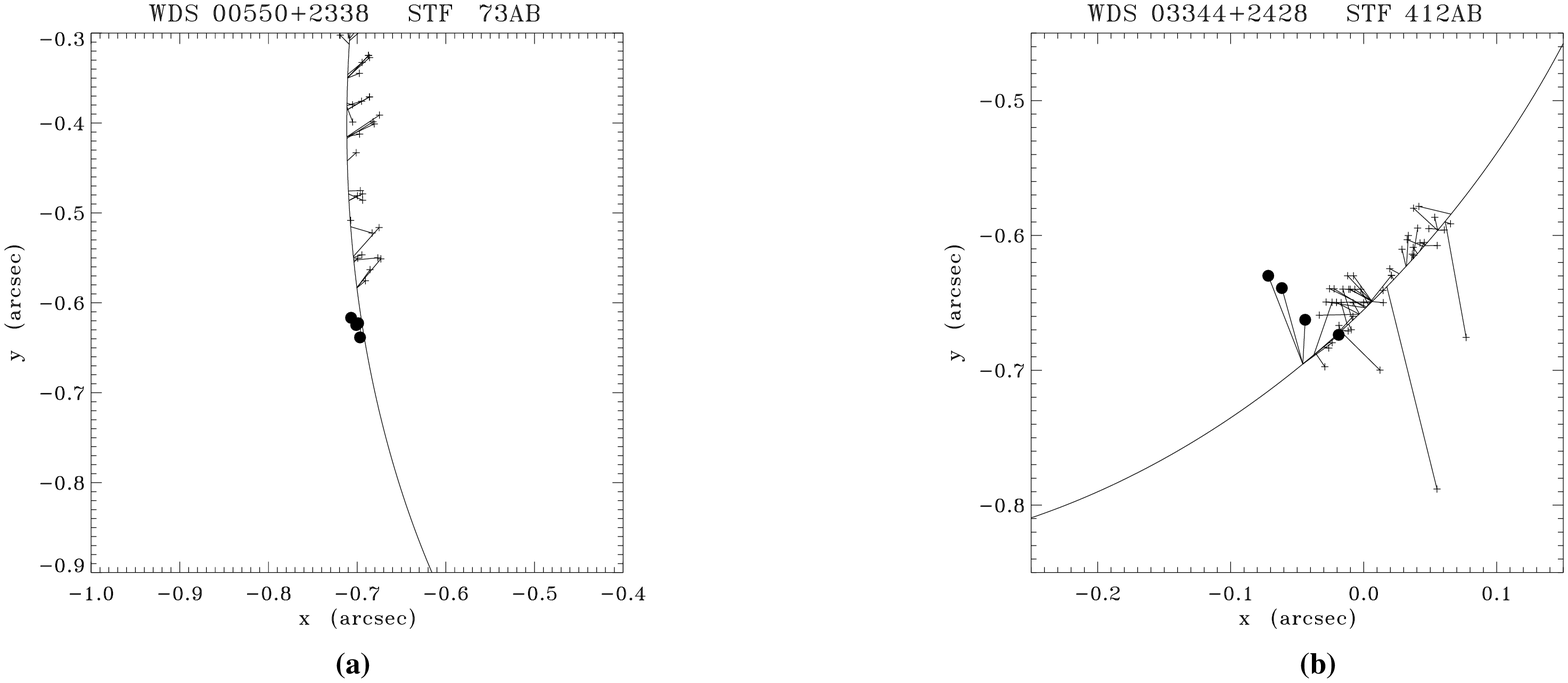}}
\caption{Comparison of astrometric results using different point source calibrations.
Point sources generated from observations of \object{ADS 3711} and \object{ADS 755} were 
used in both cases. The plot symbols and orbital trajectories are the
same as in Fig.~\ref{orbits}.
(a) WDS 00550+2338 = STF 73AB = \object{ADS 755}.
(b) WDS 03344+2428 = STF 412AB = \object{ADS 2616}.}
\label{scatter}
\end{figure*}

In Fig.~\ref{orbits} we compare the obtained results with those from other
observers and the predicted orbit for each object. In general, our
measure orbit offsets are within the global scatter of all other positions.
Those that are farthest from the orbital ephemeris positions correspond,
again, to observations performed at low elevation. The point source calibrator
in all cases was generated from a high signal-to-noise observation of \object{ADS 755}.

Assuming no major systematic errors, the total uncertainty for the measures in
Table 1 can be estimated by combining the uncertainty generated from 
night-to-night scatter when using the same point source and the variation in
the result obtained by using different point source calibrators. Errors
resulting from the fitting procedure are not considered since they have been
found to be an small fraction ($\sim5$\%) of the overall uncertainty of a given
measure of $\rho$, $\theta$ and $\Delta_m$. Although the data set here does not
permit definitive uncertainty estimates due to the small sample of objects
observed, we can nonetheless make first order estimates of these quantities.
Firstly, we obtain the night-to-night scatter ($\sigma^{nn}$) by computing the
standard deviations of the two objects in Table~\ref{measures} with the largest
number of observations, and averaging those two quantities. Secondly, we
estimate point source error ($\sigma^{ps}$) by making use of the values in
Table~\ref{comparison}, which are also displayed in Fig.~\ref{scatter}. This
table includes ($\rho$,$\theta$,$\Delta_m$) results obtained when using
different point source calibrators for one single speckle sequence. The average
of the two rows designated as $\sigma$ represents an estimate of the point
source error for one observation ($\sigma^{ps}_{1}$). Whereas,
($\rho$,$\theta$,$\Delta_m$) derived in Table~\ref{measures} proceed from 5
consecutive speckle pattern sequences. As a result, to obtain a $\sigma^{ps}$
fully comparable with $\sigma^{nn}$, $\sigma^{ps}_{1}$ has been divided by
$\sqrt{n-1}$, $n=5$. 

Finally, assuming statistical independence, we obtain the following expected
uncertainties in each coordinate by adding $\sigma^{ps}$ in quadrature with
$\sigma^{nn}$:

\indent Position angle: $\sigma_{\theta} = 1.5^{\circ}$,\\
\indent Separation: $\sigma_{\rho} = 0.017^{\prime \prime}$,\\
\indent Magnitude difference: $\sigma_{\Delta m} = 0.34$ mag.

The separation number is very similar to the result in Douglass et al.
(\cite{douglass99}) (U.S. Naval Observatory obtained speckle results with
$\sigma_{\rho} = 0.018^{\prime \prime}$ using 66-cm telescope). However,
$\sigma_{\theta}$ is higher in our case (Douglass et al. obtained
$0.57^{\circ}$ for a $1^{\prime \prime}$ separation, although $1^{\circ}$ is a
typical uncertainty in well-calibrated speckle work). $\sigma_{\Delta m}$ is
probably large because of the small window used and self-calibration technique
limitations.

\begin{figure}
\resizebox{\hsize}{!}{\includegraphics{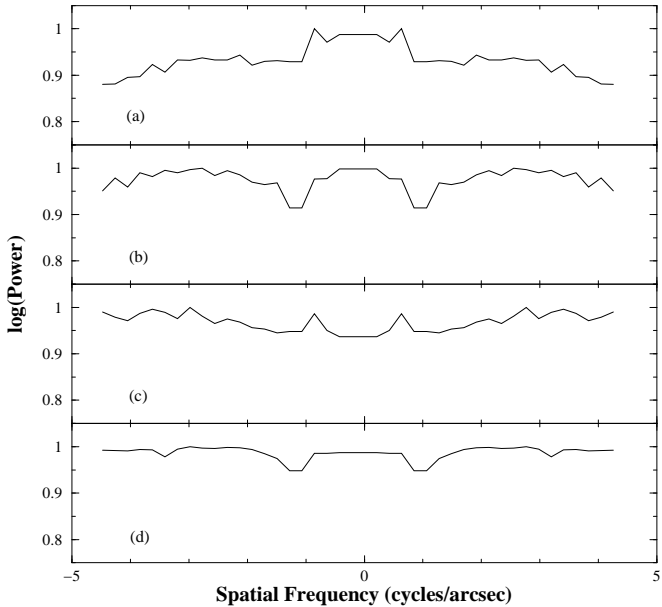}}
\caption{Comparison of 1-D point source power spectrum of \object{ADS 755} on
the 5th night of observation with respect to: (a) \object{ADS 16386} on 5th night,
(b) \object{ADS 16386} on 6th night, (c) \object{ADS 755} on 2nd night and
(d) \object{ADS 755} on 4th night.}
\label{comp_pwr}
\end{figure}

As stated above, the point source from \object{ADS 755} was used for the
analysis of all objects. To find the degree of validity of this assumption, and
to determine how significant the change in atmospheric conditions is, we have
divided the point source 1-D power spectrum of \object{ADS 755} by those from
\object{ADS 16836} and \object{ADS 755}, obtained on different nights. Ideally,
the resulting curves should be constant and equal to unity for all
frequencies.  As shown in Fig.~\ref{comp_pwr}, the curves appear to be quite
flat over the whole frequency domain. Only marginal residuals in the range of
seeing wings are visible for the two upper plots. Those are due to region of
the seeing peak not being considered when the power spectrum fits are performed.
The information in Fig.~\ref{comp_pwr} is complementary to what is shown in
Table~\ref{comparison}.

%______________________________________________________________
\section{Discussion}

\subsection{New CCDs improvements} \label{improv}

The performance of the fast drift scanning technique depends on camera readout
rate, which directly fixes the minimum dead time attainable. If this is
excessive, streaking could handicap the scientific usefulness of the data. In
general, the readout rate of a camera will depend on its digitization level and
data transfer rate. The digitization rate is fixed by CCD microcontroller
design and digitization depth. The data transfer rate is specified by the port
architecture being used between the camera and computer. 

In the case of the camera used in current paper, its readout rate of
30~kpix~s$^{-1}$ can be considered as moderately low compared with CCDs currently
in the market. The fact that it is controlled through parallel port interface
fairly limits the final readout rate. This is not surprising, since SBIG-ST8 was
designed for general CCD stare imaging purposes, where a long download time was
not the main concern. 

However, in the last few years, technologies directly related to CCD
performance have undergone significant developments. Taking into account only
those which apply to full frame CCDs, i.e. the type used most in astronomy, we
can consider the following advances:

$\bullet$ readout noise, which has been continuously dropping in all kind of
cameras. Still on the edge of the professional market, the recently available
L3Vision technology is able to offer cameras with sub-electron readout noise
(e2v Technologies, \cite{e2v02}).

$\bullet$ multi output CCD, which increases the frame rate by dividing the data
stream to be readout into several channels.

$\bullet$ new data transfer interfaces, which have noticeably increased the 
throughput in comparison to parallel port (see Table~\ref{transfer}). Apogee
(\cite{apogee03}), SBIG (\cite{sbig02}) and Ethernaude (\cite{ethernaude01})
constitute recent examples of this improvement, with the application of USB 2.0
and Ethernet interfaces to CCDs in the high-end amateur market. These
kinds of initiative can supply readout rates typically 10 to 30 times faster
than that offered by our port-parallel camera.

Therefore, the benefits to the fast drift scanning technique from all CCD
improvements above are straightforward. On one hand, lower readout noise will
increase the SNR of the specklegrams. On the other hand, a faster readout rate
will certainly decrease dead time and, as a result, low-level streaking between
speckle frames would be effectively reduced.

\begin{table}
\begin{center}
\caption[]{Data transfer rate for different port architectures.}
\label{transfer}
\begin{tabular}{lc}
\hline \hline
Type & Data transfer rate \\
 & (Mbit s$^{-1}$) \\
\hline
Serial & 0.115 \\
Parallel Port EPP/ECP & 0.5-1 \\
Firewire & 200 \\
USB 2.0 & 480 \\
Ethernet & 10/100/1000 \\
\hline
\end{tabular}
\end{center}
\end{table}

\subsection{Limitations of the self-calibration technique} \label{limit}

The self-calibration method used here cannot be used in all situations. Indeed,
the principal limitation is due to the zenith angle. As the zenith angle
increases, the dispersion of the atmosphere elongates the speckles so that the
speckle transfer function is no longer radially symmetric, and therefore the
point source estimate generated is not an appropriate representation of the
speckle transfer function at high zenith angles. This in turn can affect the
relative astrometry and photometry derived from such data. 

In considering differential photometry, one would expect that this is more
sensitive to calibration effects than the astrometric results, since the
process of deriving the magnitude difference amounts to estimating the fringe
depth in the Fourier plane.  If one uses a symmetric point spread function
estimate to deconvolve an asymmetric binary power spectrum, the fringe depth
can be severely affected while the fringe spacing and orientation would remain
essentially the same.

It is also quite likely that in the case of a faint binary star, it is probably
better to use a brighter binary to obtain the one-dimensional trace simply due
to signal-to-noise considerations.

%______________________________________________________________
\section{Summary and final remarks}

A new approach to performing CCD-based speckle imaging has been presented.
Data obtained by those means have enough quality to give real scientific
results, as shown for objects observed for this paper.

In addition, a new approach for calibrating the power spectrum analysis has
been introduced. It does not require point source observations, which gives a
more effective use of observation time. Some limitations have been observed for
this method for zenith angles above  $ 30^{\circ}$ related to atmospheric
dispersion. These conclusions can gain even more importance for the case of
large telescopes. On the one hand, as they have the highest observing time
pressure, self-calibration techniques would obviate point sources observations.
On the other hand, if conveniently equipped with Risley prisms, they could be
used to observe objects at low elevations without serious effects on the shape
of speckles due to atmospheric dispersion. Thus, self-calibration would
presumably not be limited by elevation.

CCDs, far from being specialized detectors, are very common among
instrumentation available in most astronomical observatories. The fast drift
scanning enables low budget professional and high-end amateur observatories,
which routinely use full-frame CCDs for stare imaging, to perform CCD speckle
imaging as well. The performance of this technique will be significantly higher
with new faster and less noisy cameras which are becoming available in the CCD
market. 

%______________________________________________________________
\begin{acknowledgements} This work was supported in part by the DGICYT
Ministerio de Ciencia y Tecnolog\'{\i}a (Spain) under grant no. AYA2001-3092.
O. Fors was supported by a fellowship from DGESIC Ministerio de Educaci\'{o}n,
Cultura i Deportes (Spain), ref. AP97~38107939. Authors thank Observatorio
Astronomico Nacional for facilities made available at Calar Alto. We would like
to express our gratitude here to Christoph Flohr for making available his
program SCAN.
\end{acknowledgements}

%______________________________________________________________

\end{document}